\documentclass{iopart}

\usepackage[comma,numbers,sort&compress]{natbib}
\usepackage{iopams}
\usepackage{setstack}
\usepackage{amsfonts}
\usepackage{amssymb}
\usepackage{multirow}
\usepackage{color}
\usepackage{colortbl}
\usepackage{setspace}

\usepackage{epsfig}

\def\newblock{\hskip .11em plus .33em minus .07em}

\def\apj{Astrophys. J.}
\def\apjl{Astrophys. J. Lett.}

\def\aap{Astron. Astrophys. }

\def\mnras{Mon. Not. Roy. Astron. Soc. }

\def\prl{Phys. Rev. Lett.}
\def\prd{Phys. Rev. D.}

\def\apss{Astrophys. Space Sci.}

\begin{document}

\title[Probing the SN Mechanism with Gravitational Waves]
{Probing the Core-Collapse Supernova Mechanism with Gravitational Waves}

\author{Christian D Ott}
\ead{cott@tapir.caltech.edu}

\address{\scriptsize 
  Theoretical Astrophysics, Mailcode 350-17,\\
           California Institute of Technology,
           Pasadena, California 91125, USA \\
  and \\ 
  Niels Bohr International Academy, Niels Bohr Institute,\\
  Copenhagen, Denmark 
  \\
  and \\
  Center for Computation and Technology,
  Louisiana State University,\\ Baton Rouge, LA, USA}

%%%%%%%%%%%%%%%%%%%%%%%%%%%%%%%%%%%%%%%%%%%%%%%%%%%%%%%%%%%%%%%%%%%%%%%%
%%%%%%%%%%%%%%%%%%%%%%%%%%%%%%%%%%%%%%%%%%%%%%%%%%%%%%%%%%%%%%%%%%%%%%%%
%%%%%%%%%%%%%%%%%%%%%%%%%%%%%%%%%%%%%%%%%%%%%%%%%%%%%%%%%%%%%%%%%%%%%%%%
\begin{abstract}
The mechanism of core-collapse supernova explosions must draw on the
energy provided by gravitational collapse and transfer the necessary
fraction to the kinetic and internal energy of the ejecta. Despite
many decades of concerted theoretical effort, the detailed mechanism
of core-collapse supernova explosions is still unknown, but
indications are strong that multi-D processes lie at its heart. This
opens up the possibility of probing the supernova mechanism with
gravitational waves, carrying direct dynamical information from the
supernova engine deep inside a dying massive star.  I present a
concise overview of the physics and primary multi-D dynamics in
neutrino-driven, magnetorotational, and acoustically-driven
core-collapse supernova explosion scenarios.  Discussing and
contrasting estimates for the gravitational-wave emission
characteristics of these mechanisms, I argue that their
gravitational-wave signatures are clearly distinct and that
the observation (or non-observation) of gravitational waves
from a nearby core-collapse event could put strong constraints
on the supernova mechanism.
\end{abstract}

\hspace{2.55cm}{\scriptsize Submitted to the GWDAW13 special issue of CQG on May 17, 2009.}

%%%%%%%%%%%%%%%%%%%%%%%%%%%%%%%%%%%%%%%%%%%%%%%%%%%%%%%%%%%%%%%%%%%%%%%%
%%%%%%%%%%%%%%%%%%%%%%%%%%%%%%%%%%%%%%%%%%%%%%%%%%%%%%%%%%%%%%%%%%%%%%%%
%%%%%%%%%%%%%%%%%%%%%%%%%%%%%%%%%%%%%%%%%%%%%%%%%%%%%%%%%%%%%%%%%%%%%%%%

\section{Introduction}
\label{section:intro}

In their 1934 paper \cite{baade:34b}, Baade and Zwicky suggested that
supernovae (SNe) are caused by the release of gravitational energy in
the ``collapse of ordinary stars to neutron stars.'' Now, seventy-five
years after this initial proposition and after much observational and
theoretical effort, Baade's and Zwicky's statement still holds and is
the fundamental paradigm of core-collapse SN theory.

Massive stars ($\sim$$8-10\,M_\odot \lesssim M \lesssim 100\,M_\odot$)
live fast nuclear-burning lives, consuming their nuclear fuel in a
cosmic blink of only $\sim100\,\mathrm{million} - 1\,\mathrm{million}$
years ($\tau / \tau_\odot \approx (M/M_\odot)^{-2.5}$). When nuclear
burning ceases in the core, a massive star has evolved to a red or
blue supergiant (radius $\sim \mathrm{few} \times
10^{13}\,\mathrm{cm}$ or $\sim 0.1 - 1 \times 10^{12}\,\mathrm{cm}$,
respectively) and is comprised of a central, compact
Chandrasekhar-mass electron-degenerate iron core ($R \sim
1500\,\mathrm{km}$) that is surrounded by an onion-skin structure of
shells of material with successively lower mean atomic weight.
Supported only by increasingly relativistically-degenerate electrons,
the iron core eventually exceeds its effective Chandrasekhar mass.
Catastrophic gravitational collapse sets in and is accelerated by
electron capture on nuclei and free protons as well as
photodissociation of iron-group nuclei. In a matter of a few hundred
milliseconds, the inner core is compressed from densities of $10^9 -
10^{10}\,\mathrm{g\,cm}^{-3}$ to nuclear density ($\sim 2.6 \times
10^{14}\,\mathrm{g\,cm}^{-3}$) where the nuclear equation of state
(EOS) stiffens and stabilizes the inner core. The latter overshoots
its new equilibrium, then rebounds into the still collapsing outer
core.  This \emph{core bounce} results in the formation of a
hydrodynamic shock at the interface of inner and outer core. The shock
propagates outward in radius and mass through the supersonically
infalling outer core material, leaving behind the unshocked $\sim
0.5-0.7\,M_\odot$ inner core which forms the core of the hot
protoneutron star (PNS). The PNS rapidly gains mass through accretion
($\dot{M}$ is of the order of $1\,M_\odot\,\mathrm{s}^{-1}$) and
radiates away its gravitational energy ($\sim 300\,\mathrm{B}$ for a
$1.4\,M_\odot$ NS; $1\, \mathrm{[B]ethe} = 10^{51}\,\mathrm{erg}$) in
neutrinos as it evolves to a cold NS on a timescale of tens of
seconds.

In early theory and SN models, a \emph{prompt} hydrodynamic explosion
would follow shock creation \cite{bethe:90}, removing the stellar
envelope and providing the fantastic optical display of a SN with
total kinetic, internal and electromagnetic energy of
$\sim 1\,\mathrm{B}$\footnote{Inferred core-collapse SN energies fall in
the range of $0.1-10\,\mathrm{B}$ \cite{hamuy:03}, but energies
around $1\,\mathrm{B}$ are most frequently seen.}.

More advanced theory and simulations with more complete and accurate
treatment of the EOS, neutrino physics and transport show that the
prompt shock fails: Photodissociation of infalling heavy nuclei into
nucleons (at a cost of $\sim 8.8\,\mathrm{MeV}\,\mathrm{baryon}^{-1}$
for iron-group nuclei) and neutrinos that stream away from the
optically-thin postshock region sap the shock's energy. The shock
slows down quickly after its launch, stalls, and turns into an
accretion shock at a radius of $100 - 200\,\mathrm{km}$. For a SN
explosion to occur, there must operate a \emph{mechanism} that, in
some fashion, transfers and deposits a fraction of the tremendous
gravitational energy of collapse into the immediate postshock region,
reviving the shock and endowing it with sufficient kinetic energy to
make a core-collapse SN.

\emph{What is the core-collapse SN mechanism?} This is the fundamental
question and primary unsolved problem of core-collapse SN
theory. Finding observational evidence for the SN mechanism by
classical astronomical means from radio to $\gamma$ wavelengths is
difficult, since all pre-explosion dynamics takes place deep inside
the stellar core, completely shrouded from view in the electromagnetic
(EM) spectrum.  Hence, EM observations yield only \emph{secondary}
observables, such as progenitor type and mass, explosion morphology 
and energy, ejecta composition, neutron star properties and
birth kick. Primary, direct ``live'' information from deep inside
the supernova engine is carried only by neutrinos and gravitational
waves (GWs) which both can propagate from their emission sites to
observers on Earth virtually without interaction with intervening
matter.

GWs have not yet been observed directly, but an international network
of interferometric detectors (LIGO, VIRGO, GEO, TAMA) is taking
data with sufficient sensitivity to observe GWs from a nearby galactic
core-collapse SN (e.g., \cite{ott:09,ligoburst:09}).  Neutrinos, on
the other hand, were detected from SN 1987A, confirming the basic
theory of stellar collapse and neutrino emission.  However, in part
due to the small number of detected neutrinos, this observation left
unanswered the question of the SN mechanism.

In this contribution to the proceedings of the 13th Gravitational-Wave
Data Analysis Workshop (GWDAW13), I delineate the current set of
proposed core-collapse SN mechanisms. I discuss the underlying physics
of each mechanism and limit my focus to the dominant multi-D processes
active and leading to GW emission in a given mechanism. Furthermore, I
argue that the GW signatures of the various SN mechanisms are distinct
and that the observation or non-observation of GWs by current and/or
future advanced or third-generation GW observatories from a nearby SN
can put significant constraints on the explosion mechanism. This idea
was put forth first in \cite{ott:09} and is elaborated here.  In
\sref{section:numech}, I discuss the neutrino mechanism and its GW
signature. This mechanism is currently favored as the explanation for
core-collapse SN explosions in garden-variety slowly or nonrotating
massive stars.  Section~\ref{section:mhdmech} is devoted to the
magnetorotational mechanism and its GW emission. This mechanism may be
active in rapidly rotating massive stars. In \sref{section:acmech}, I
discuss the recently proposed acoustic mechanism of
Burrows~et~al.~\cite{burrows:06,burrows:07a} which relies on the
excitation of PNS pulsations and their damping via acoustic waves that
steepen to shocks and efficiently deposit energy in the postshock
region.  In section~\ref{section:sum},  I summarize and 
draw conclusions.

\section{The Neutrino Mechanism}
\label{section:numech}

The \emph{neutrino-driven mechanism} (or simply, \emph{neutrino
  mechanism}), is founded on the notion that if only a small fraction of
the $E_\mathrm{grav} \sim 300\,\mathrm{B}$ of gravitational energy
liberated in collapse and emitted in neutrinos and antineutrinos of
all flavors was deposited in the stellar mantle, it could be driven to
explosion. The neutrino mechanism, based on the charged-current
absorption of $\nu_e$ and $\bar{\nu}_e$ on neutrons and protons,
respectively, was first proposed by Colgate \& White~\cite{colgate:66}
and Arnett~\cite{arnett:66} and was initially envisioned to be
efficient and direct, that is, expected to drive the shock out
before it has time to stall. However, simulations with improved
numerics and input microphysics showed that neutrino energy deposition
(``neutrino heating'') is actually rather inefficient and the shock
always stalls first and may be revived only on a timescale of hundreds
of milliseconds in the \emph{delayed neutrino mechanism} proposed by
Bethe \& Wilson~\cite{bethewilson:85}.  The amount of energy to be
delivered by the mechanism, $E_\mathrm{mech}$, is set by the sum of
asymptotic explosion energy of $\sim 1\,\mathrm{B}$ and the
gravitational binding energy of the mantle (also of the order of
$1\,\mathrm{B}$) that must first be overcome. Hence, $E_\mathrm{mech}
/ E_\mathrm{grav} \sim 1\%$ is the fraction of the total energy that
must be transferred to the mantle by neutrinos to power an explosion.
This $1\%$ must be provided to the
shock within a short, $\lesssim 1-2\,\mathrm{s}$ time frame after
bounce in order for the onset of explosion to occur before accretion
has pushed the PNS over its maximum mass and for the
SN to leave behind a NS mass consistent with observations. Typical $\nu_e$
and $\bar{\nu}_e$ luminosities in the postbounce phase are $L_{\nu_e}
\approx L_{\bar{\nu}_e} \approx 10\, \mathrm{B}\,\mathrm{s}^{-1}$ and,
hence, assuming that the explosion must start within $\sim
1\,\mathrm{s}$, the necessary \emph{heating efficiency} is of the
order of $\sim 10\%$. 

As compelling and promising the neutrino mechanism may seem, the most
advanced simulations predict its failure in ordinary massive stars
with masses $\gtrsim 10\,M_\odot$ when spherical symmetry (1D) is
assumed and fluid motions are restricted to the radial direction
\cite{ramppjanka:00,thompson:03,liebendoerfer:05}. Only the
lowest-mass massive stars with $7\, M_\odot \lesssim M \lesssim
10\,M_\odot$ (S-AGB stars that end their nuclear burning lives with
O-Ne cores) may explode, though sub-energetically, by the 1D neutrino
mechanism alone \cite{kitaura:06,burrows:07c}.  In multi-D,
neutrino-driven convection \cite{herant:94,bhf:95,jankamueller:96} in
combination with the standing-accretion-shock instability (SASI)
(e.g., \cite{scheck:08,iwakami:07,fernandez:08} and references
therein) can enhance the heating efficiency
\cite{buras:06b,murphy:08}.  Yet, the presently most complete
axisymmetric (2D) simulations still fall short of producing robust
$1$-$\mathrm{B}$ neutrino-driven explosions
\cite{buras:06b,marek:09,burrows:06,ott:08} and hope for the neutrino
mechanisms rests on the additional degrees of freedom provided by 3D
dynamics to facilitate robust explosions in nature and in future
detailed 3D simulations~\cite{fryerwarren:04,murphy:08,iwakami:07}.

\begin{figure}
\centering
\includegraphics[width=0.45\linewidth]{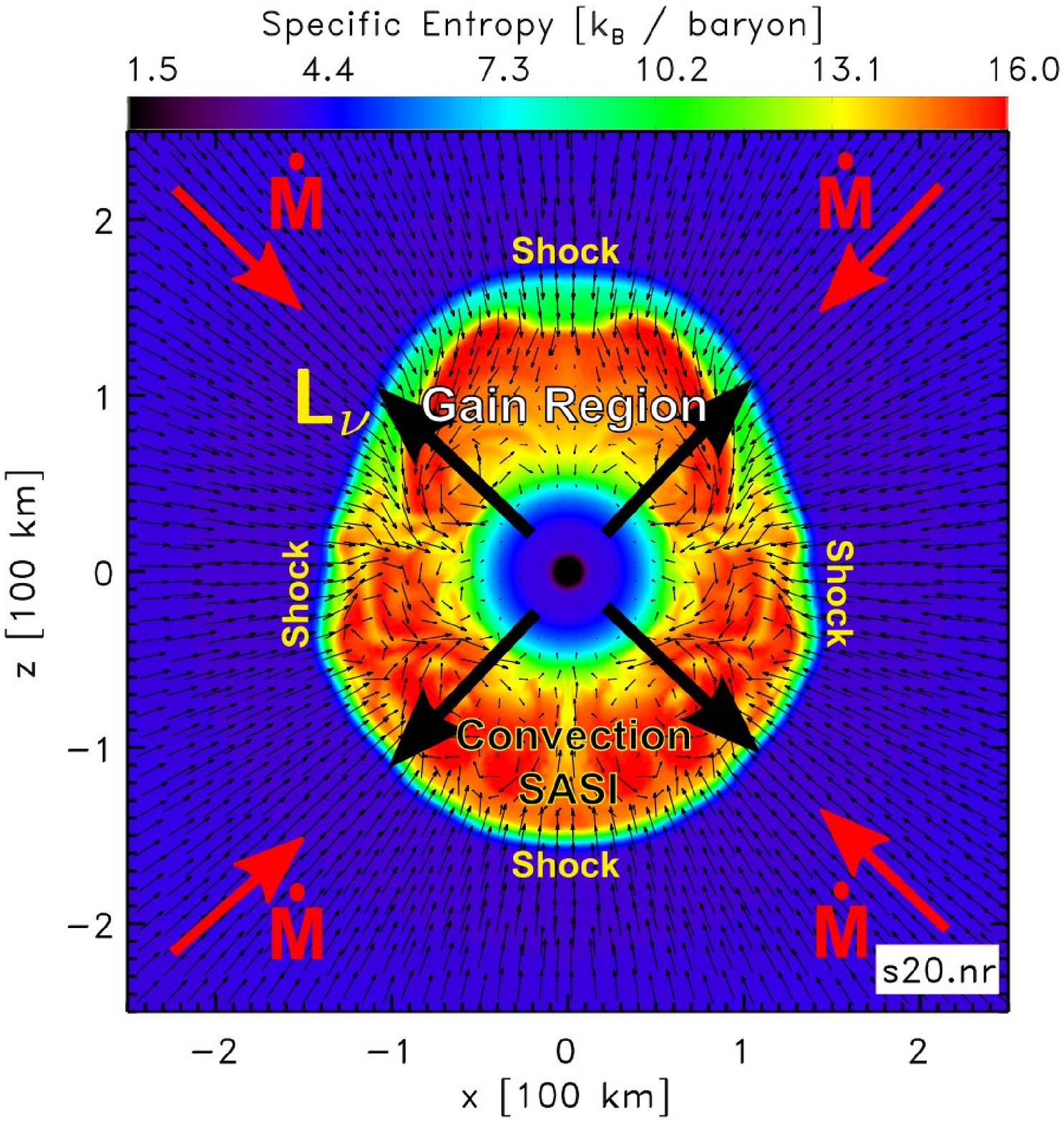}
\includegraphics[width=0.45\linewidth]{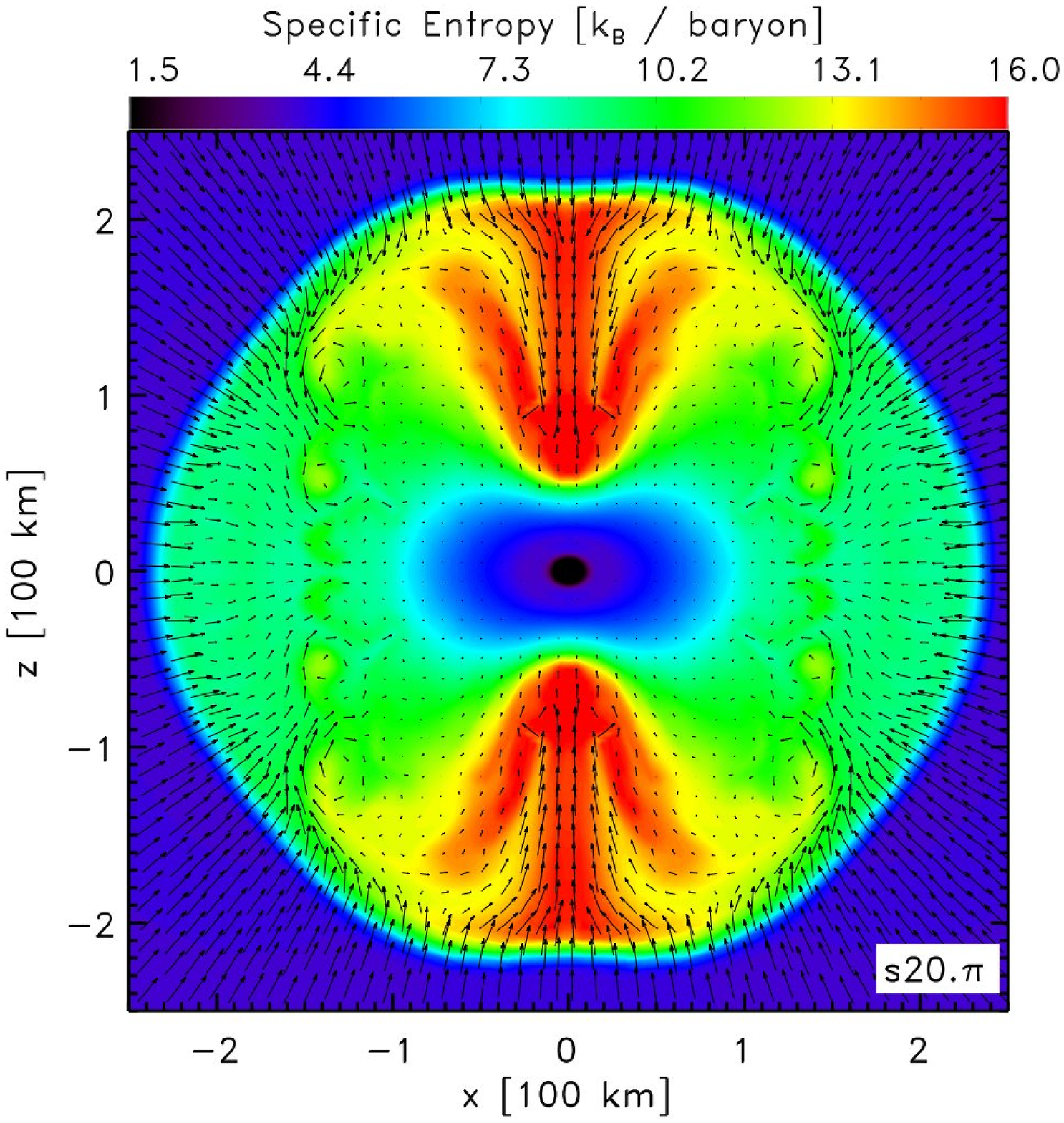}
\caption{Postbounce supernova cores of the nonrotating model s20.nr
  (left panel) and the rapidly-spinning model s20.$\pi$ (right panel)
  of the angle-dependent neutrino-radiation-hydrodynamics simulations
  of Ott~et~al.~\cite{ott:08} at $\sim 215\,\mathrm{ms}$ after
  bounce. Shown are the specific entropy distributions and velocity
  vectors are superposed to visualize flow patterns. Key aspects of
  the postbounce situation are highlighted in model s20.nr in the left
  panel.}
\label{fig:sn_cartoon}
\end{figure}

Figure~\ref{fig:sn_cartoon} displays color maps of the specific
entropy distribution in snapshots taken at $\sim 215\,\mathrm{ms}$
after bounce of the simulations of Ott~et~al.~\cite{ott:08} that
started with a $20$-$M_\odot$ presupernova model and were carried out
without rotation (model s20.nr) and with rapid precollapse rotation
(model s20.$\pi$, $\Omega_0 = \pi\,\mathrm{rad\,s}^{-1}$). Simulation
details are explained at length in \cite{ott:08}. The nonrotating
model is depicted in the left panel and the various features of the
postbounce situation are highlighted by text and arrows. Neutrinos are
trapped in the dense PNS core, but leak out from the
neutrinosphere\footnote{The neutrinosphere is defined as the location
  at which the neutrino optical depth $\tau_\nu$ is $2/3$. The radius
  $R_\nu$ of the neutrino sphere is strongly dependent on neutrino
  energy.}  at the PNS surface. Neutrino cooling ($Q^{-}_{\nu} \propto
T^6$, e.g., \cite{janka:01}) dominates over neutrino heating
($Q^{+}_\nu \propto L_\nu r^{-2} \langle \epsilon_\nu^2\rangle$,
\cite{janka:01}) interior to the gain radius of $80-100\,\mathrm{km}$
denoting the location beyond which $Q^{+}_\nu > Q^{-}_{\nu}$. The net
neutrino heating is strongest near the inner boundary of the gain
region and decreases outwards. Hence, a negative entropy gradient
develops, resulting in instability to overturn and convective eddies
in the gain region are clearly visible. Also discernible is the
deformation of the shock front due to the early phase of the SASI
which drives the growth of low-mode ($\ell = {1,2}$) perturbations
in an advective-acoustic cycle of laterally propagating sound waves and
advection of pressure perturbations~(e.g., \cite{fernandez:08} and references
therein).

\subsection{Convection and SASI in Nonrotating and Rotating SN Cores}

In the core-collapse SN context, convective instability can generally
be driven by negative gradients in entropy $s$ or lepton fraction $Y_l
= Y_e + Y_\nu$. 
Rotation or more specifically, a
positive specific angular momentum gradient, can stabilize the flow
against overturn. This is expressed in the Solberg-H\o iland condition
for instability. In the equatorial plane and neglecting
lateral gradients, this condition is given by
\begin{equation}
N^2 + \frac{1}{r^3} \frac{d}{dr} j^2 < 0\,\,.
\label{eq:sh}
\end{equation}
Here, $j = \Omega r^2$ is the specific angular momentum and
$N^2$ is the Brunt-V\"ais\"ala frequency, in the SN context 
\cite{fh:00,thompson:05},
\begin{equation}
N^2 = \frac{g}{p \gamma}\left(\frac{\partial p}{\partial s}\bigg|_{\rho, Y_l} 
\frac{ds}{dr} 
+ \frac{\partial p}{\partial Y_l}\bigg|_{\rho,s} \frac{d Y_l}{dr}\right)\,\,,
\end{equation}
where $g$ is the gravitational acceleration, $\gamma = d\ln
p/d\ln\rho|_s$, and $\partial p/\partial s$ and $\partial p/\partial
Y_l$ are generally positive.  In the right panel of
\fref{fig:sn_cartoon}, the entropy distribution and the superposed
velocity vector field of the rapidly spinning (PNS period $\sim
2\,\mathrm{ms}$) oblate postbounce SN core of model s20.$\pi$ is shown
at $\sim 215\,\mathrm{ms}$ after bounce.  The large positive specific
angular momentum gradient in its equatorial regions makes the
left-hand side of eq.~\eref{eq:sh} positive and convective overturn is
confined to regions near the axis and large equatorial radii where the
$j$-gradient flattens off.  The shock in model s20.$\pi$ is deformed
to oblate shape and its average radius is considerably greater than at
the same time in the nonrotating model, but a top-bottom asymmetry
typical for the SASI is absent. According to the results of
Ott~et~al.~\cite{ott:08}, rotation not only inhibits convection, but
also delays and alters the growth of the axisymmetric SASI significantly.
As pointed out by \cite{yamasaki:08,iwakami:08}, the situation may
be different in 3D where a spiral-type azimuthal SASI may develop in
rapidly rotating SN cores.

In nonrotating and slowly rotating core-collapse SNe, convection
generically occurs in three distinct ways: (1) Prompt convection,
driven by the negative entropy gradient left behind by the stalled
shock and strongly dependent on the seed perturbations present in the
core \cite{marek:09,ott:09}, (2) as lepton-gradient driven convection
in the PNS (e.g., \cite{dessart:06a,keil:96,buras:06b}), and (3) as the
already mentioned neutrino-driven convection in the gain region behind
the shock (e.g., \cite{herant:94,bhf:95,jankamueller:96,buras:06b,burrows:06}).

\subsection{GW Emission}
\label{section:convgw}

\begin{figure}
\centering
\hspace*{-.75cm}\includegraphics[width=0.54\linewidth]{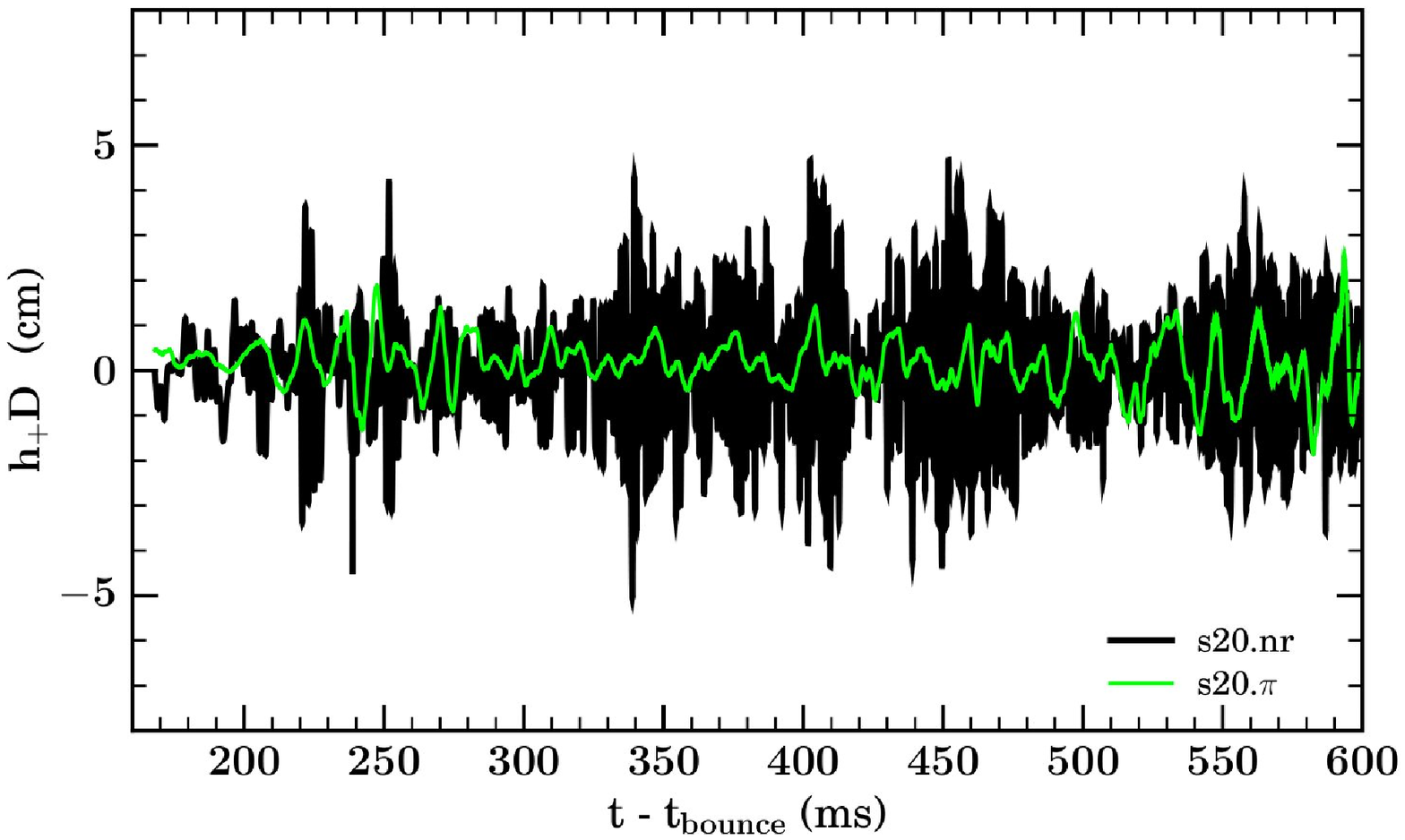}
\includegraphics[width=0.54\linewidth]{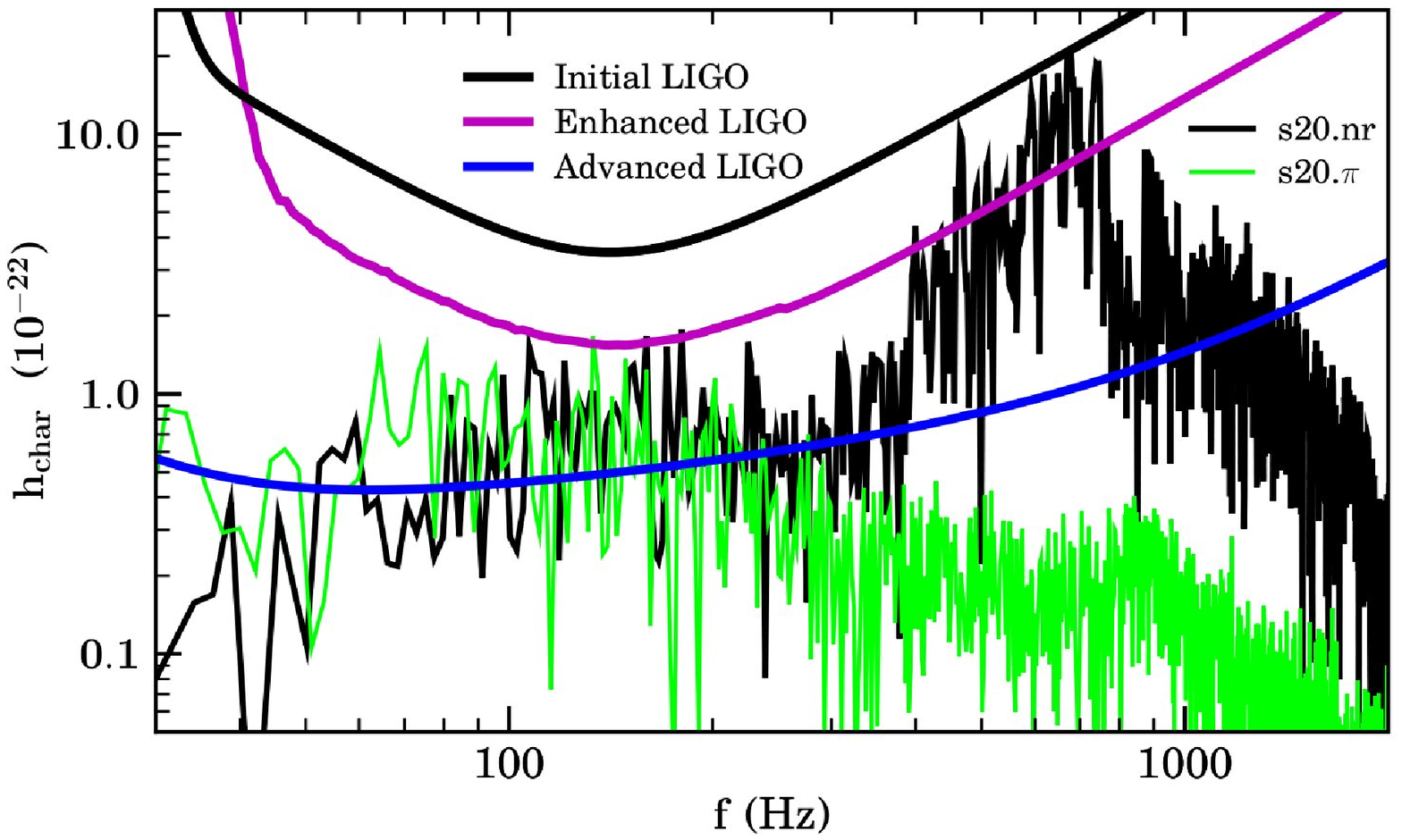}\hspace*{-0.75cm}
\caption{{\bf Left:} GW signal $h_+$ (scaled by the source distance
  $D$ and in units of cm) of convection in the PNS and neutrino-driven
  convection and SASI in the postshock region in the axisymmetric
  postbounce models s20.nr (nonrotating, black lines) and s20.$\pi$
  (rapidly rotating, green lines) of \cite{ott:08}. Rapid rotation
  effectively damps convection and modifies the SASI, leading to a
  lower-amplitude and more slowly varying GW signal. {\bf Right:}
  Characteristic strain spectra $h_\mathrm{char}$ \cite{flanhughes:98}
  of the same GW signals assumed to be emitted at $D =
  10\,\mathrm{kpc}$ and contrasted with theoretical initial, enhanced,
  and advanced LIGO design noise curves. Note that, in practice,  initial
  LIGO's low-frequency sensitivity is identical to that of enhanced LIGO~\cite{rana:09}.}
\label{fig:convgw}
\end{figure}

In nonrotating or slowly rotating core-collapse SNe and in the absence
of strong PNS pulsations (discussed in
\sref{section:acmech}), the dominant multi-D dynamics and, hence, the
primary emission process of GWs is associated with convective overturn
and the SASI. In addition, and also related to the asymmetry 
due to convection and SASI, anisotropic emission of neutrinos
adds a slowly-varying component to the GW signal with characteristic
frequencies of the order of $10\,\mathrm{Hz}$. Recent quantitative
estimates of the latter can be found in
\cite{marek:09b,ott:09,kotake:09}.  Here, I focus on the
higher-frequency, more readily-detectable GW signal from fluid motions
and present in the left-hand panel of \fref{fig:convgw} the GW signal
from the postbounce simulations of Ott~et~al.~\cite{ott:08} (models
s20.nr and s20.$\pi$ as discussed above) that were run with
full 2D angle-dependent neutrino-radiation hydrodynamics. The GW
signals include the contributions from PNS convection, neutrino-driven
convection and SASI. Scaled to a fiducial galactic source distance of
$10\,\mathrm{kpc}$, the peak amplitudes correspond to a dimensionless
GW strains of $\sim 1.3\times10^{-22}$ and are reached by the
nonrotating model s20.nr. Convection in the rapidly rotating model is
strongly inhibited, both in the PNS and in the
postshock region. This is reflected in its low-amplitude, slowly-varying
GW signal.

In the right panel of \fref{fig:convgw}, I plot the characteristic
gravitational wave strain of the two postbounce SN models, defined by
Flanagan \& Hughes~\cite{flanhughes:98} as $h_\mathrm{char}(f) =
{D}^{-1} \sqrt{ 2 {\pi^{-2}} {G}{c^{-3}} dE_{\mathrm{GW}}
  {df}^{-1}}\,, $ where $D$ is the distance to the source (set to
$10\,\mathrm{kpc}$) and $d E_\mathrm{GW}/ df$ is the GW spectral
energy density. In order to contrast the $h_\mathrm{char}$ spectra
with detector sensitivity, I also plot the design sensitivity
$h_\mathrm{rms}$ of initial LIGO~\cite{ligo}, enhanced
LIGO~\cite{oshaughnessy:09}, and advanced LIGO (in burst mode)
\cite{shoemaker:06}. According to the simulation data of
Ott~et~al., in a nonrotating or slowly-rotating SN, most of the energy
in GWs is emitted in a broad frequency range of
$300-1000\,\mathrm{Hz}$ (this is in rough agreement with
\cite{marek:09}) by a combination of the high-$f$, small-scale PNS
convection and downflow plumes of material caused by the SASI that
penetrate to small radii and are rapidly decelerated at the PNS
surface \cite{marek:09,ott:09}. The emission due to convection in the
rapidly rotating model is much weaker, lacks the high-$f$ component
and occurs primarily at frequencies of $\sim 10 - 200\,\mathrm{Hz}$.

It is important to note two aspects: (1) The nature of the GW emission
by convection and SASI is stochastic~\cite{ott:09,kotake:09} and the
emitted GWs will be of uncorrelated, essentially random
polarization. This is a consequence of the turbulent, chaotic emission
dynamics that depend in a non-deterministic fashion on local
conditions and can vary greatly from one SN to the next.  (2) The
detectability of the GWs from convection and SASI (i.e., the magnitude
of $h_\mathrm{char}$ in the right panel of \fref{fig:convgw}) depends
crucially on the duration of emission. The SASI and neutrino-driven
convection shut off once an explosion has fully developed or a BH has
formed.  Hence, it may last $\sim 200 \,\mathrm{ms} - 3\,\mathrm{s}$,
the upper limit depending on the onset of explosion, the accretion
rate set by the progenitor and the EOS-dependent mass-limit for BH
formation. PNS convection would be shut off by BH formation, but if an
energetic explosion occurs, it is likely to continue for many seconds
until the PNS has reached neutrino-less $\beta$-equilibrium. See
\cite{ott:09} for scalings of the emitted GW energy with the duration
of the convective signal.

\section{The Magnetorotational Mechanism}
\label{section:mhdmech}

Most ($\gtrsim 99\%$) core-collapse SN progenitors are expected to be
rather slowly rotating and have central periods of tens to hundreds of
seconds \cite{heger:05,ott:06spin}. However, there exists a variety of
progenitor channels that could yield iron cores with precollapse
periods of the order of seconds~(e.g.,
\cite{woosley:06,fryer:05}). Conservation of angular momentum in
collapse leads to a spin-up by a factor of $\sim
1000$~\cite{ott:06spin}. Hence, an iron core with $P_0 \sim
2\,\mathrm{s}$ results in a PNS with $P \sim 2\,\mathrm{ms}$ and a
rotational energy of the rigidly-rotating PNS and of the strongly
differentially rotating postshock layer of a $\mathrm{few} \times
10\,\mathrm{B}$~\cite{ott:06spin,burrows:07b}.  A fraction of this
energy would be sufficient to power an energetic core-collapse SN
provided there is an efficient mediating mechanism for converting
rotational energy into linear kinetic energy of the ejecta.

Theory and early simulations
\cite{leblanc:70,meier:76,bisno:76,symbalisty:84} have shown that
magnetorotational processes may constitute such a mechanism and can
drive collimated outflows, leading to energetic jet-driven bipolar
explosions~\cite{wheeler:02}.  Recently, this \emph{magnetorotational}
mechanism has received much attention (e.g., \cite{akiyama:03,
  shibata:06,burrows:07b,sawai:08,takiwaki:09} and references therein)
owing to increasing observational evidence for collimated jets in the
context of gamma-ray bursts (GRBs) and hypernovae (hyper-energetic
$\gtrsim 10$-$\mathrm{B}$ SNe) and the association of stellar collapse
with the long-soft class of GRBs~(e.g., \cite{wb:06}).

Recent
simulations~\cite{burrows:07b,dessart:08a,sawai:08,shibata:06,takiwaki:09}
suggest that magnetic fields of the order of $10^{15}\,\mathrm{G}$
with strong toroidal components are required to yield the necessary
magnetic stresses to drive a strong bipolar explosion.  Rapidly
rotating progenitors may have rather strongly magnetized cores, with
toroidal fields of $\sim 10^{9} - 10^{11}\,\mathrm{G}$ (the poloidal
components are $1-2$ orders of magnitude weaker)
\cite{heger:05,woosley:06}.  Flux compression in collapse scales
$\propto (\rho/\rho_0)^{2/3}$ \cite{burrows:07b,shibata:06} and can
lead to an amplification of both toroidal and poloidal components by
up to a factor of $1000$.  Further amplification is provided for the
toroidal component by rotational winding of poloidal into toroidal
field ($\Omega$-dynamo, e.g., \cite{cerda:07,burrows:07b,shibata:06},
a linear process) and by the exponentially-growing magneto-rotational
instability~(MRI, e.g.,
\cite{akiyama:03,shibata:06,obergaulinger:09,burrows:07b} and
references therein) that has the potential to provide
dynamically-relevant saturation field strengths of the order of
$10^{15}\,\mathrm{G}$ \cite{obergaulinger:09}. Both, $\Omega$-dynamo
and MRI operate on rotational shear $(d\Omega/d\ln r)$ and convert
energy stored in differential rotation\footnote{The lowest-energy
  state of a rotating fluid at fixed total angular momentum is uniform
  rotation. Hence, the difference in energy between uniform and
  differential rotation at fixed total angular momentum can be
  considered as ``free energy of differential rotation.''} into
magnetic field.  Strong differential rotation in the outer SN core is
a generic feature of rotating core
collapse~\cite{ott:06spin,akiyama:05}.

The to-date most microphysically complete MHD core-collapse
simulations (though in a Newtonian framework) were carried out by
Burrows and collaborators~\cite{burrows:07b,dessart:08a}. Under the
provision that the MRI operates as expected~\cite{obergaulinger:09}
and the postbounce rotational energy is of the order of $\sim
10\,\mathrm{B}$ (implying $P_0 \lesssim 4 \,\mathrm{s}$), these
authors find robust and powerful bipolar magnetorotational explosions
that reach hypernova energies, leave behind a protomagnetar and could
be the stage-setting precursor to a subsequent GRB
\cite{burrows:07b,dessart:08a}.

\begin{figure}
\centering
\hspace*{-.75cm}\includegraphics[width=0.54\linewidth]{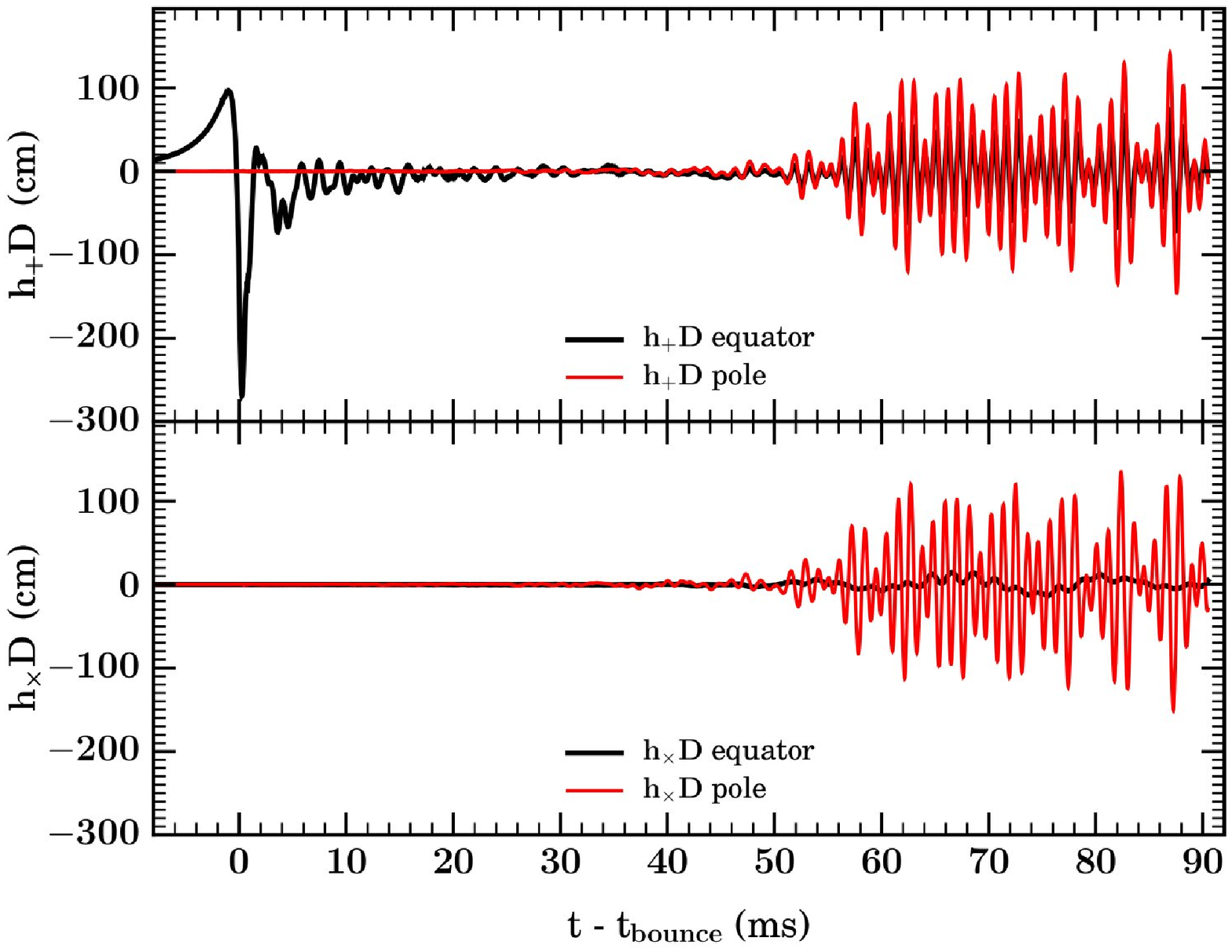}
\includegraphics[width=0.49\linewidth]{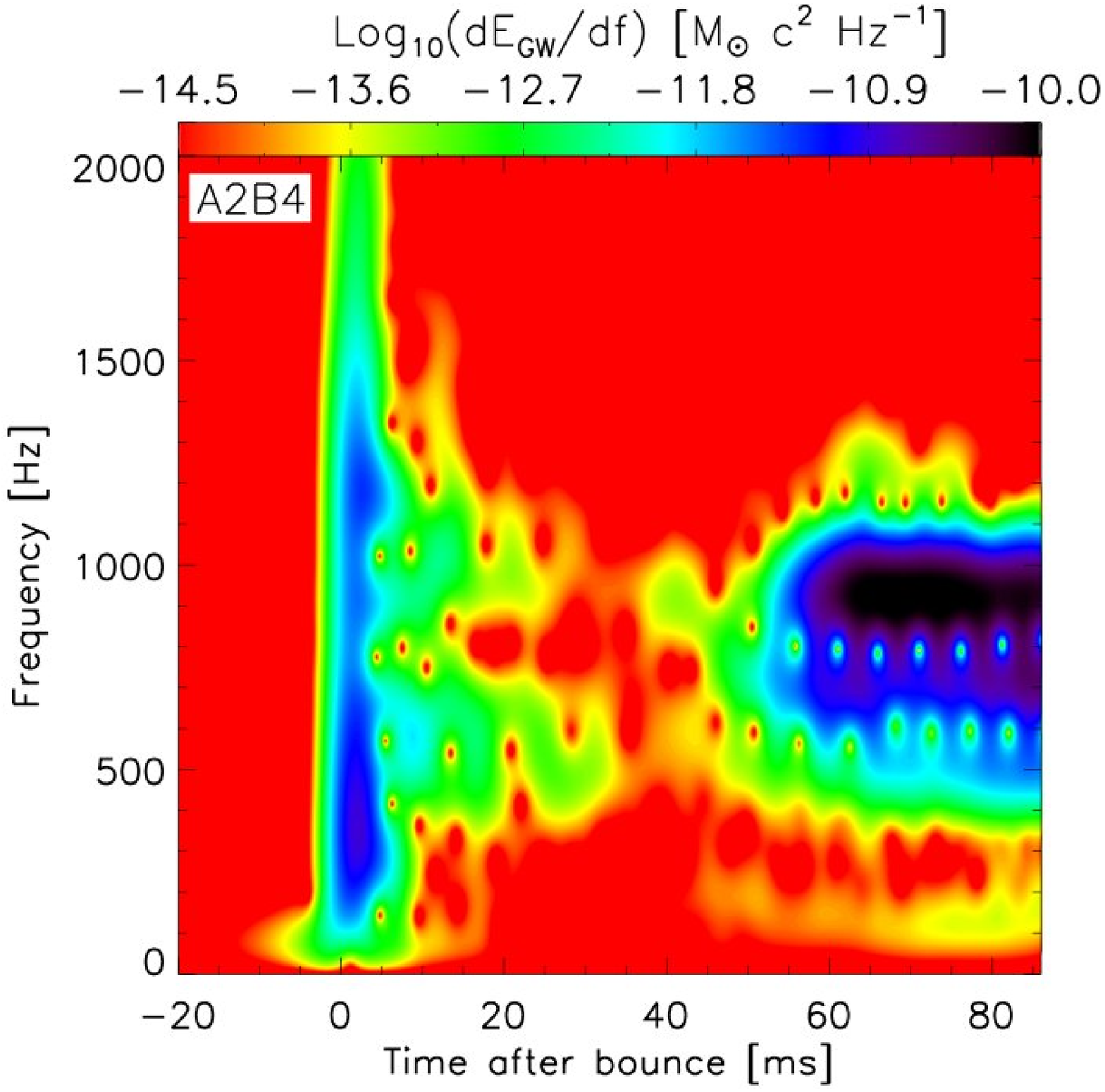}\hspace*{-0.75cm}
\caption{{\bf Left:} GW signal (rescaled by distance $D$ and in units
  of cm) emitted by rotating core collapse, bounce and postbounce
  nonaxisymmetric dynamics in model s20A2B4 of
  \cite{ott:07prl,ott:07cqg}. Shown are the $h_+$ (top) and $h_\times$
  (bottom) polarizations as seen by observers situated in the
  equatorial plane (red lines) and along the polar axis (black
  lines). Note that the GWs are linearly polarized during the
  axisymmetric collapse and bounce phase and become predominantly
  elliptically polarized only tens of milliseconds after bounce.  This
  panel is a variant of the right panel of figure~4 in
  \cite{ott:09}. {\bf Right:} Time-frequency (TF) analysis of the
  spectral energy density of the GWs emitted by model s20A2B4. The TF
  analysis was carried out with a $2-\mathrm{ms}$ Gaussian window that
  was shifted over the data in steps of $0.2\,\mathrm{ms}$. Both the
  strong burst at core bounce and the energetic late postbounce
  emission leave clear and clearly separated marks in the TF diagram.}
\label{fig:nonaxi}
\end{figure}

\subsection{GW Emission}

Without any reasonable doubt one can assume that rapid rotation is the
primary multi-D component of the magnetorotational mechanism.  Rapid
rotation leads to a strong burst of GWs at core bounce, associated with the huge
accelerations acting on the centrifugally deformed inner core. 
This GW emission process is the most extensively studied in
the core-collapse context (see, e.g., \cite{ott:09} for a review) and
the currently best estimates of the GW signal come from 2D and 3D GR
simulations that include a microphysical EOS and deleptonization during
collapse~\cite{ott:07prl,ott:07cqg,dimmelmeier:07,dimmelmeier:08}. These
studies have demonstrated that the GW signal from bounce is of generic
morphology, exhibiting a small pre-bounce rise, a large negative peak
roughly coincident with bounce and a subsequent ringdown signal
emitted as the PNS settles to its new quasi-hydrostatic
configuration. Comparing 2D and 3D simulations,
\cite{ott:07prl,ott:07cqg} demonstrated that even
extremely rapidly rotating cores stay axisymmetric through
bounce, which constrains the GW emission to linear polarization
(emission only in $h_+$). According to the results of
\cite{dimmelmeier:08}, a core collapsing with a precollapse period
$P_0 \sim 2\,\mathrm{s}$ ($P_0 \sim 4\,\mathrm{s}$) located at
$10\,\mathrm{kpc}$ emits a peak GW amplitude of
$|h_\mathrm{max}|\sim1\times10^{-21}$ ($|h_\mathrm{max}|\sim
5\times10^{-21}$) and a total energy in GWs of the order of a
$\mathrm{few} \times 10^{-8}\,M_\odot c^2$ with most of the emission
being concentrated at $500-800\,\mathrm{Hz}$, extending to lower
frequencies with increasing rotation.

Although realistic PNSs are unlikely to undergo the classical
dynamical MacLaurin-type nonaxisymmetric instability at rotation rates
$T/|W|\gtrsim 27\%$~\cite{dimmelmeier:08}, nonaxisymmetric dynamics
may still develop via a low-$T/|W|$ corotation
instability~\cite{watts:05}. For equilibrium NS models, this
was first discovered by~\cite{centrella:01} and has since been found
to occur also in more realistic postbounce SN settings
~\cite{rotinst:05,scheidegger:08,ott:07prl,ott:07cqg}.

In the left panel of \fref{fig:nonaxi}, I plot the gravitational wave
polarizations $h_+$ (top) and $h_\times$ (bottom) as seen by
equatorial (black lines) and polar observers (red lines) emitted by
model s20A2B4 in the $3+1\,\mathrm{GR}$ framework of
Ott~et~al.~\cite{ott:07prl,ott:07cqg}. This model used a $20$-$M_\odot$
progenitor with am iron core set up to spin roughly
uniformly at a period of $\sim 1\,\mathrm{s}$. The purely axisymmetric
($\ell = 2, m = 0$ in terms of spherical harmonics) bounce signal is
followed by a primarily axisymmetric ringdown.  Nonaxisymmetric
dynamics develops in the postbounce phase and becomes relevant some
$40\,\mathrm{ms}$ after bounce as indicated by the rise of the GW
signal emitted along the poles ($\ell = 2, m = 2$) due to the
quadrupole components of the nonaxisymmetric dynamics.  The right
panel of the same figure displays a time-frequency analysis of the GW
spectral energy density. The quick change in the waveform at bounce
leads to power in a broad range of frequencies, but most of the energy
is emitted around $350-400\,\mathrm{Hz}$ in this rather rapidly
rotating core. The nonaxsiymmetric component kicks in at higher
frequencies around $\sim 900-950\,\mathrm{Hz}$ (with secondary,
modulating components at $\sim 700\,\mathrm{Hz}$ and $\sim
500\,\mathrm{Hz}$) and, despite emitting lower-amplitude GWs, is
significantly more energetic than the bounce signal, emitting roughly
$E_\mathrm{GW} \sim 1.6\times10^{-7}\,M_\odot c^2$ until the
simulation was terminated. The nonaxisymmetric dynamics could
last longer, possibly for hundreds of milliseconds~\cite{ott:07cqg,ott:09},
but may be in competition with the MRI, an aspect that remains yet to be
investigated. 

\section{The Acoustic Mechanism}
\label{section:acmech}

The \emph{acoustic mechanism}, proposed in a series of papers by
Burrows and collaborators~\cite{burrows:06,burrows:07a,ott:06prl}
(using 2D radiation-hydrodynamic simulations), is based on the
excitation of PNS core pulsations (primarily $\ell = \{1,2\}$ $g$-modes)
by turbulence and by accretion downstreams through the unstable and
highly-deformed stalled shock in the non-linear phase of the SASI.  In
the simulations of Burrows~et~al., fueled by the gravitational energy
of anisotropic accretion, the PNS pulsations ramp up to non-linear
amplitudes over many hundreds of milliseconds and damp via the emission of
strong sound waves. These travel down the steep density gradient
present in the postshock region, thus steepen to shocks and very
efficiently deposit their energy behind and in the shock. Explosions
in the acoustic mechanism are robust, but set in only after $\sim
1\,\mathrm{s}$ after bounce and tend to have energies on the lower
side of what is observed.

\begin{figure}
\centering
\hspace*{-.75cm}\includegraphics[width=0.54\linewidth]{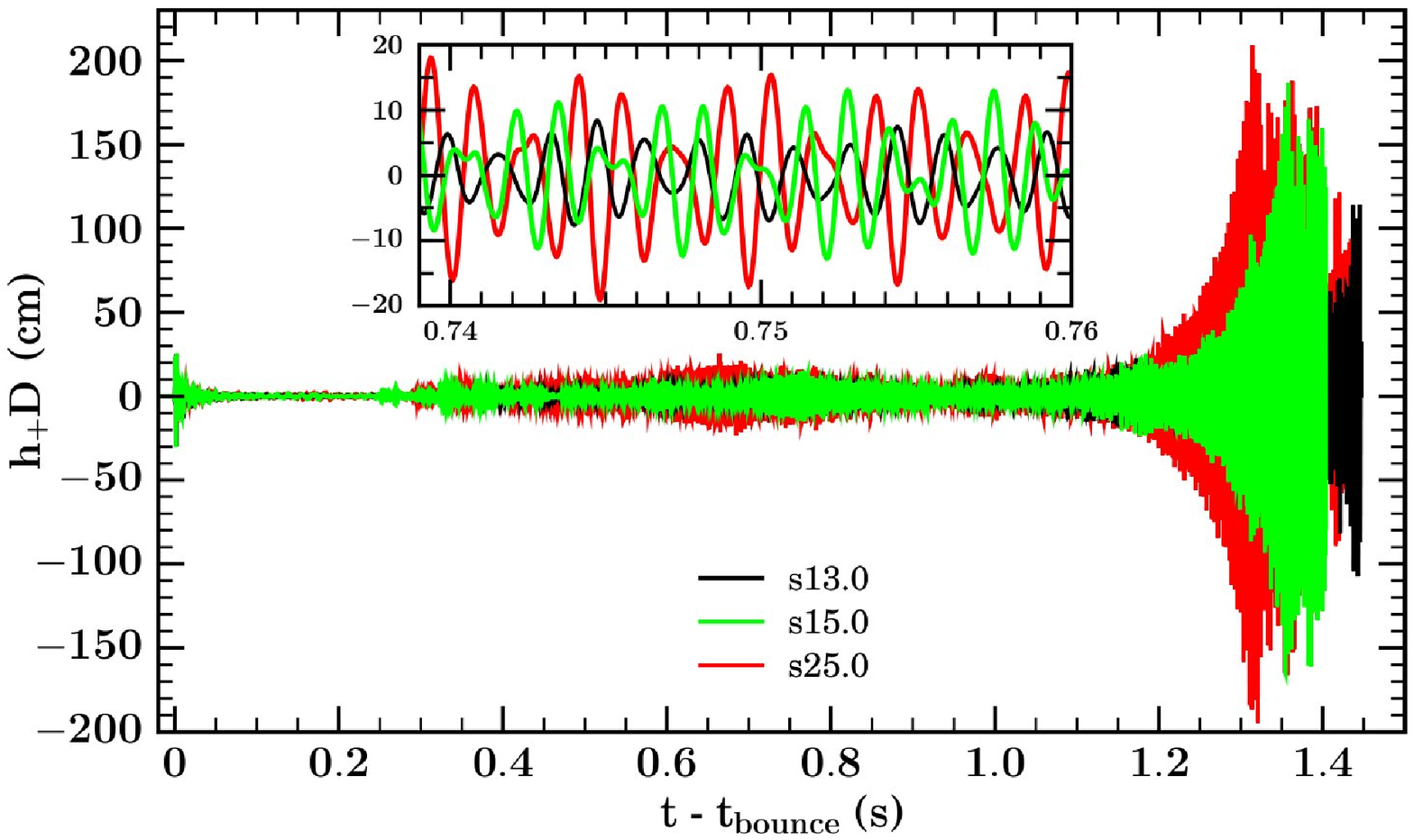}
\includegraphics[width=0.54\linewidth]{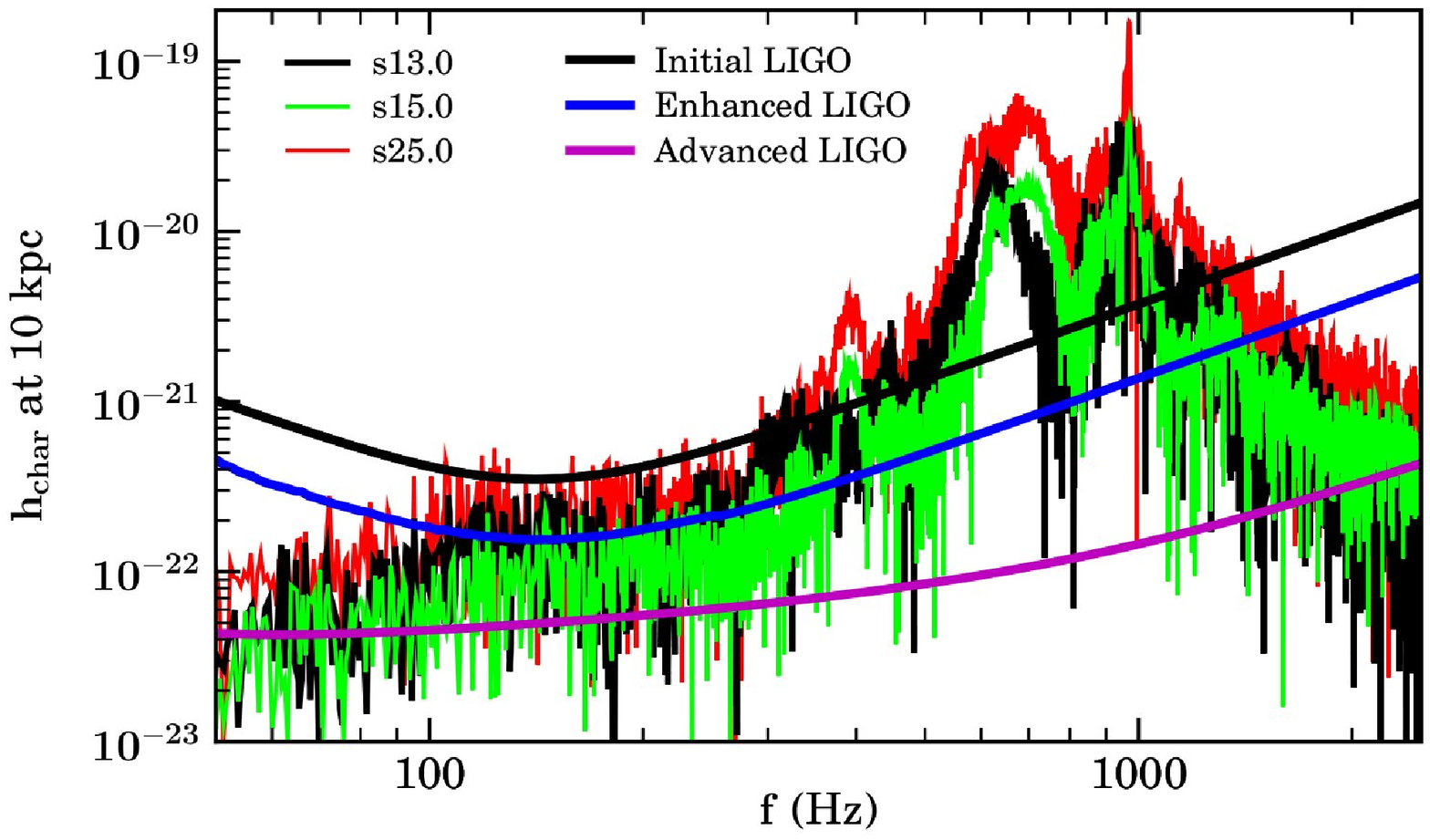}\hspace*{-0.75cm}
\caption{{\bf Left:} GW signals (rescaled by distance $D$ and in units
  of cm) emitted in a set of representative nonrotating models
  from~\cite{burrows:07a} that explode via the acoustic mechanism. At
  early times, the GW signal contains a burst associated with prompt
  convection at bounce and a longer-term broad-band low-amplitude
  contribution from PNS convection, neutrino-driven convection, and
  the SASI. At later times ($t \gtrsim 300-600\,\mathrm{ms}$ after
  bounce), the GW emission becomes dominated by PNS core pulsations,
  leading to a narrowband signal (a zoomed-in view is shown in the
  inset plot). Around the time of the onset of explosion the largest
  GW amplitudes are reached. {\bf Right:} Characteristic GW
  strain spectra $h_\mathrm{char}$ \cite{flanhughes:98} (at
  $10\,\mathrm{kpc}$) of the same models contrasted with initial,
  enhanced and advanced LIGO design noise curves.  }
\label{fig:gmodegw}
\end{figure}

There are multiple caveats associated with the acoustic mechanism: (1)
It is so-far unconfirmed, but also not yet ruled out by other SN 
groups (e.g., \cite{marek:09}). (2) Perturbation theory
suggests that the PNS mode amplitudes may be limited by a parametric
instability involving high-order modes that are not presently resolved
in numerical simulations~\cite{weinberg:08}. (3) If the neutrino
mechanism is effective, it will probably explode the star before the
PNS pulsations have time to grow to large amplitudes.  (4) As pointed
out in \sref{section:numech}, rapid rotation has a stabilizing effect
on convection and SASI and, hence, is likely to also inhibit the growth
of PNS pulsations. In addition, rapidly rotating SNe are likely to explode
by the magnetorotational mechanism much before PNS pulsations can reach
large amplitudes. (5) It is not clear how the relaxation of axisymmetry
would affect the power and spatial character of the PNS pulsations. 3D
simulations have yet to be performed.

\subsection{GW Emission}

The early postbounce GW emission associated with the acoustic
mechanism is similar to that of the neutrino mechanism discussed in
\sref{section:convgw} and comes primarily from convection, the SASI
and SASI downflow plumes that are decelerated at the PNS core's outer
edge. The latter are drivers of the PNS pulsations whose quadrupole
components dominate the GW emission in the later postbounce phase. They
are the most characteristic dynamical feature of the acoustic mechanism.

In the left panel of \fref{fig:gmodegw}, I plot sample GW signals
extracted from the simulations of Burrows~et~al.~\cite{burrows:07a} (a
more detailed discussion of these models was presented in
\cite{ott:09}) for a variety of progenitors. Present is a burst of
prompt convection around core bounce, followed by a phase of
neutrino-driven convection with little SASI. Around $\sim
300-600\,\mathrm{ms}$ after bounce, the non-linear SASI regime is reached,
downflow plumes reach the PNS core and the core pulsations begin to
grow and emit GWs at the frequencies of their quadrupole components
(see inset plot) that are set by the core's structure. While the GW
emission in these models at intermediate times stems from $\ell = 2$
harmonics of the then dominant $\ell = 1$ mode, an $\ell = 2$
eigenmode at higher frequencies appears dominant at times $\gtrsim 1 -
1.2\,\mathrm{s}$, explaining the increase of the GW amplitudes 
around the onset of explosion. 

The PNS core pulsations of the acoustic mechanism are the strongest GW
emission mechanism proposed in the stellar collapse context. Predicted
strain amplitudes range up to $h_+ \sim 10^{-20}$ (at
$10\,\mathrm{kpc}$, \cite{ott:09}) and total GW energies are
$E_\mathrm{GW} \sim 10^{-7}\,M_\odot c^2$ and above, emitted in a
narrow frequency band set by the PNS pulsation frequencies and their
temporal evolutions. The right panel of \fref{fig:gmodegw},
contrasting $h_\mathrm{char}$ spectra with detector design
sensitivities, demonstrates that a galactic SN exploding via the
acoustic mechanism would be difficult to miss even with
first-generation detectors.

\section{Discussion and Conclusions}
\label{section:sum}

\begin{figure}
\centering
\hspace*{-.75cm}\includegraphics[width=0.4\linewidth]{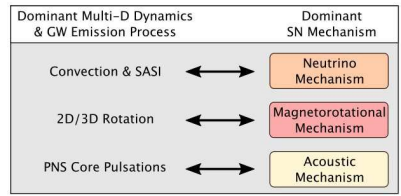}
\caption{Scheme summarizing the associations of particular multi-D
  processes and characteristic GW emissions with the individual
  candidate core-collapse SN mechanisms.}
\label{fig:mech}
\end{figure}

Each of the three SN mechanism that I consider in this article
has an unique multi-D dynamical process intimately linked
with it. The acoustic mechanism, while involving neutrino-driven
convection and SASI, is fundamentally reliant on non-linear PNS core
pulsations to drive the acoustic SN engine. In the magnetorotational
mechanism, on the other hand, rapid rotation is the key ingredient,
damping convection and SASI, and facilitating B-field amplification
and a bipolar jet-driven explosion. Finally, the neutrino mechanism's
key multi-D aspect is convection and SASI. While PNS core pulsations
could certainly be excited, they would not have time to grow to large
amplitudes if the neutrino mechanism was effective and launched an
early explosion.

Now, as delineated in the previous sections and portrayed by figures
\ref{fig:convgw}$-$\ref{fig:gmodegw}, the three distinct multi-D
processes dominant in the three SN mechanisms have very different GW
emission characteristics: (1) Convection and SASI of the neutrino
mechanism lead to a low-amplitude, broadband ``stochastic''
signal. (2) The rapid rotation necessary for the magnetorotational
mechanism is reflected in GWs by a strong axisymmetric burst at core
bounce and a possible prolonged narrowband postbounce GW signal from
nonaxisymmetric dynamics. (3) The strong PNS core pulsations of the
acoustic mechanism -- if they obtain -- emit copiously in narrow
intervals about their characteristic frequencies.

With the above information a simple picture can be drawn
(\fref{fig:mech}), relating a particular SN mechanism with a
particular flavor of multi-D dynamics and an unique GW
signature. Hence, with an observation of GWs from a core-collapse SN,
the first strong observational constraints could be put on the SN
mechanism. The magnetorotational and acoustic mechanisms should be
clearly visible throughout the Milky Way with initial and enhanced
interferometric detector technology while advanced and probably
third-generation observatories (e.g., ET \cite{ET}) may be needed to
constrain the SN mechanism for core-collapse events at
$3-5\,\mathrm{Mpc}$ out to which the integrated SN rate is $\sim
0.5\,\mathrm{yr}^{-1}$~\cite{ando:05,ott:09}. Due to the
low-amplitude, broadband nature of its GW emission, the neutrino
mechanism is the most difficult to detect in GWs, but a non-detection
of GWs from a galactic event with initial and enhanced detectors would
quite clearly rule out the other two candidate mechanisms.

The way in which I have portrayed the candidate SN mechanisms and
their GW signatures in this article is idealized: I have focussed only
on the primary multi-D dynamics in each mechanism and its
associated GW emission, neglecting contributions from secondary and,
in particular, low-frequency components, such as GWs from anisotropic
neutrino emission, magnetic stresses, or explosion asymmetries (see
\cite{ott:09} for a detailed discussion). I have also neglected the
possibility of the combined action of multiple mechanisms. For
example, an explosion could be driven by a combination of neutrino
heating and MHD processes in a moderately rapidly rotating SN.  Yet,
even in such a case, the dominant dynamical multi-D component would
also dominate the GW emission and the simple scheme displayed in
\fref{fig:mech} is essentially preserved.

Further theoretical and physically more complete and accurate
computational work is necessary to gain a better understanding of the
various possible SN mechanisms and of the details of their GW
signatures. But, as I have attempted to show in this article, even
with current knowledge, the GW signal of a galactic SN today may very
likely provide strong hints for one mechanism and/or
smoking-gun evidence against another.

\section*{Acknowledgements}
It is a pleasure to thank E.~O'Connor, E.~Abdikamalov, R.~Adhikari,
A.~Burrows, L.~Cadonati, L.~Dessart, H.~Dimmelmeier, I.~S. Heng,
H.-T.~Janka, E.~Katsavounidis, S.~Klimenko, K.~Kotake, A.~Marek,
C.~Meakin, J.~Murphy, R.~O'Shaughnessy, B.~Owen, C.~Pethick,
E.~S.~Phinney, E.~Schnetter, U.~Sperhake, and K.~Thorne for helpful
and stimulating discussions.  This work was supported by a Sherman
Fairchild postdoctoral fellowship at Caltech and by an Otto Hahn Prize
awarded to the author by the Max Planck Society. Results presented in
this article were obtained through computations on the NSF Teragrid
under grant TG-MCA02N014, on machines of the Louisiana Optical Network
Initiative under grant LONI\_NUMREL03, and at the National Energy
Research Scientific Computing Center (NERSC), which is supported by
the Office of Science of the US Department of Energy under contract
DE-AC03-76SF00098.

{\scriptsize \setlength{\parskip}{0.1cm} 

}


\begin{thebibliography}{71}
\expandafter\ifx\csname natexlab\endcsname\relax\def\natexlab#1{#1}\fi
\expandafter\ifx\csname url\endcsname\relax
  \def\url#1{{\tt #1}}\fi

\bibitem[{Baade} and {Zwicky}(1934)]{baade:34b}
W.~{Baade} and F.~{Zwicky}.
\newblock {\em Proc. Nat. Acad. Sci.}, {\bf 20}, 259, 1934.

\bibitem[{Bethe}(1990)]{bethe:90}
H.~A. {Bethe}.
\newblock {\em Rev. Mod. Phys.}, {\bf 62}, 801, 1990.

\bibitem[{Hamuy}(2003)]{hamuy:03}
M.~{Hamuy}.
\newblock {\em \apj}, {\bf 582}, 905, 2003.

\bibitem[{Ott}(2009)]{ott:09}
C.~D {Ott}.
\newblock {\em Classical and Quantum Gravity}, {\bf 26(6)}, 063001, 2009.

\bibitem[Ligo Scientific~Collaboration(2009)]{ligoburst:09}
{{Abbot}, B.~P. et al.,} Ligo Scientific~Collaboration.
\newblock {\em arXiv:0905.0020 [gr-qc]}, 2009.

\bibitem[{Burrows} et~al.(2006){Burrows}, {Livne}, {Dessart}, {Ott}, and
  {Murphy}]{burrows:06}
A.~{Burrows}, E.~{Livne}, L.~{Dessart}, C.~D. {Ott}, and J.~{Murphy}.
\newblock {\em Astrophys. J.}, {\bf 640}, 878, 2006.

\bibitem[{Burrows} et~al.(2007{\natexlab{a}}){Burrows}, {Livne}, {Dessart},
  {Ott}, and {Murphy}]{burrows:07a}
A.~{Burrows}, E.~{Livne}, L.~{Dessart}, C.~D. {Ott}, and J.~{Murphy}.
\newblock {\em Astrophys. J.}, {\bf 655}, 416, 2007.

\bibitem[{Colgate} and {White}(1966)]{colgate:66}
S.~A. {Colgate} and R.~H. {White}.
\newblock {\em \apj}, {\bf 143}, 626, 1966.

\bibitem[{Arnett}(1966)]{arnett:66}
W.~D. {Arnett}.
\newblock {\em Canadian Journal of Physics}, {\bf 44}, 2553, 1966.

\bibitem[{Bethe} and {Wilson}(1985)]{bethewilson:85}
H.~A. {Bethe} and J.~R. {Wilson}.
\newblock {\em \apj}, {\bf 295}, 14, 1985.

\bibitem[{Rampp} and {Janka}(2000)]{ramppjanka:00}
M.~{Rampp} and H.-T. {Janka}.
\newblock {\em \apjl}, {\bf 539}, L33, 2000.

\bibitem[{Thompson} et~al.(2003){Thompson}, {Burrows}, and
  {Pinto}]{thompson:03}
T.~A. {Thompson}, A.~{Burrows}, and P.~A. {Pinto}.
\newblock {\em \apj}, {\bf 592}, 434, 2003.

\bibitem[{Liebend{\"o}rfer} et~al.(2005){Liebend{\"o}rfer}, {Rampp}, {Janka},
  and {Mezzacappa}]{liebendoerfer:05}
M.~{Liebend{\"o}rfer}, M.~{Rampp}, H.-T. {Janka}, and A.~{Mezzacappa}.
\newblock {\em \apj}, {\bf 620}, 840, 2005.

\bibitem[{Kitaura} et~al.(2006){Kitaura}, {Janka}, and
  {Hillebrandt}]{kitaura:06}
F.~S. {Kitaura}, H.-T. {Janka}, and W.~{Hillebrandt}.
\newblock {\em \aap}, {\bf 450}, 345, 2006.

\bibitem[{Burrows} et~al.(2007{\natexlab{b}}){Burrows}, {Dessart}, and
  {Livne}]{burrows:07c}
A.~{Burrows}, L.~{Dessart}, and E.~{Livne}.
\newblock {The Multi-Dimensional Character and Mechanisms of Core-Collapse
  Supernovae}.
\newblock In S.~{Immler} and R.~{McCray}, editors, {\em {AIP Conference
  Series}}, volume 937, page 370, 2007.

\bibitem[{Herant} et~al.(1994){Herant}, {Benz}, {Hix}, {Fryer}, and
  {Colgate}]{herant:94}
M.~{Herant}, W.~{Benz}, W.~R. {Hix}, C.~L. {Fryer}, and S.~A. {Colgate}.
\newblock {\em \apj}, {\bf 435}, 339, 1994.

\bibitem[{Burrows} et~al.(1995){Burrows}, {Hayes}, and {Fryxell}]{bhf:95}
A.~{Burrows}, J.~{Hayes}, and B.~A. {Fryxell}.
\newblock {\em \apj}, {\bf 450}, 830, 1995.

\bibitem[{Janka} and {M\"uller}(1996)]{jankamueller:96}
H.-T. {Janka} and E.~{M\"uller}.
\newblock {\em \aap}, {\bf 306}, 167, 1996.

\bibitem[{Scheck} et~al.(2008){Scheck}, {Janka}, {Foglizzo}, and
  {Kifonidis}]{scheck:08}
L.~{Scheck}, H.-T. {Janka}, T.~{Foglizzo}, and K.~{Kifonidis}.
\newblock {\em \aap}, {\bf 477}, 931, 2008.

\bibitem[{Iwakami} et~al.(2008{\natexlab{a}}){Iwakami}, {Kotake}, {Ohnishi},
  {Yamada}, and {Sawada}]{iwakami:07}
W.~{Iwakami}, K.~{Kotake}, N.~{Ohnishi}, S.~{Yamada}, and K.~{Sawada}.
\newblock {\em \apj}, {\bf 678}, 1207, 2008.

\bibitem[{Fern{\'a}ndez} and {Thompson}(2008)]{fernandez:08}
R.~{Fern{\'a}ndez} and C.~{Thompson}.
\newblock {\em submitted to ApJ, ArXiv:0812.4574}, 2008.

\bibitem[{Buras} et~al.(2006){Buras}, {Janka}, {Rampp}, and
  {Kifonidis}]{buras:06b}
R.~{Buras}, H.-T. {Janka}, M.~{Rampp}, and K.~{Kifonidis}.
\newblock {\em \aap}, {\bf 457}, 281, 2006.

\bibitem[{Murphy} and {Burrows}(2008)]{murphy:08}
J.~W. {Murphy} and A.~{Burrows}.
\newblock {\em \apj}, {\bf 688}, 1159, 2008.

\bibitem[{Marek} and {Janka}(2009)]{marek:09}
A.~{Marek} and H.-T. {Janka}.
\newblock {\em \apj}, {\bf 694}, 664, 2009.

\bibitem[{Ott} et~al.(2008){Ott}, {Burrows}, {Dessart}, and {Livne}]{ott:08}
C.~D. {Ott}, A.~{Burrows}, L.~{Dessart}, and E.~{Livne}.
\newblock {\em \apj}, {\bf 685}, 1069, 2008.

\bibitem[{Fryer} and {Warren}(2004)]{fryerwarren:04}
C.~L. {Fryer} and M.~S. {Warren}.
\newblock {\em \apj}, {\bf 601}, 391, 2004.

\bibitem[{Janka}(2001)]{janka:01}
H.-T. {Janka}.
\newblock {\em \aap}, {\bf 368}, 527, 2001.

\bibitem[{Fryer} and {Heger}(2000)]{fh:00}
C.~L. {Fryer} and A.~{Heger}.
\newblock {\em \apj}, {\bf 541}, 1033, 2000.

\bibitem[{Thompson} et~al.(2005){Thompson}, {Quataert}, and
  {Burrows}]{thompson:05}
T.~A. {Thompson}, E.~{Quataert}, and A.~{Burrows}.
\newblock {\em \apj}, {\bf 620}, 861, 2005.

\bibitem[{Yamasaki} and {Foglizzo}(2008)]{yamasaki:08}
T.~{Yamasaki} and T.~{Foglizzo}.
\newblock {\em \apj}, {\bf 679}, 607, 2008.

\bibitem[{Iwakami} et~al.(2008{\natexlab{b}}){Iwakami}, {Kotake}, {Ohnishi},
  {Yamada}, and {Sawada}]{iwakami:08}
W.~{Iwakami}, K.~{Kotake}, N.~{Ohnishi}, S.~{Yamada}, and K.~{Sawada}.
\newblock {\em submitted to ApJ, arXiv:0811.0651 [astro-ph]}, 2008.

\bibitem[{Dessart} et~al.(2006){Dessart}, {Burrows}, {Livne}, and
  {Ott}]{dessart:06a}
L.~{Dessart}, A.~{Burrows}, E.~{Livne}, and C.~D. {Ott}.
\newblock {\em \apj}, {\bf 645}, 534, 2006.

\bibitem[{Keil} et~al.(1996){Keil}, {Janka}, and {M\"uller}]{keil:96}
W.~{Keil}, H.-T. {Janka}, and E.~{M\"uller}.
\newblock {\em \apjl}, {\bf 473}, L111, 1996.

\bibitem[{Marek} et~al.(2009){Marek}, {Janka}, and {M{\"u}ller}]{marek:09b}
A.~{Marek}, H.-T. {Janka}, and E.~{M{\"u}ller}.
\newblock {\em \aap}, {\bf 496}, 475, 2009.

\bibitem[{Kotake} et~al.(2009){Kotake}, {Iwakami}, {Ohnishi}, and
  {Yamada}]{kotake:09}
K.~{Kotake}, W.~{Iwakami}, N.~{Ohnishi}, and S.~{Yamada}.
\newblock {\em Astrophys. J. Lett. in press, ArXiv:0904.4300 [astro-ph]}, 2009.

\bibitem[{Flanagan} and {Hughes}(1998)]{flanhughes:98}
{\'E}.~{\'E}. {Flanagan} and S.~A. {Hughes}.
\newblock {\em \prd}, {\bf 57}, 4535, 1998.

\bibitem[Adhikari(2009)]{rana:09}
R.~Adhikari.
\newblock {\em Private communiciation}, 2009.

\bibitem[lig()]{ligo}
URL \url{http://ligo.caltech.edu}.
\newblock {LIGO}.

\bibitem[O'Shaughnessy(2009)]{oshaughnessy:09}
R.~O'Shaughnessy.
\newblock {\em Private communiciation}, 2009.

\bibitem[Shoemaker(2006)]{shoemaker:06}
D.~Shoemaker.
\newblock {\em Private communiciation}, 2006.

\bibitem[{Heger} et~al.(2005){Heger}, {Woosley}, and {Spruit}]{heger:05}
A.~{Heger}, S.~E. {Woosley}, and H.~C. {Spruit}.
\newblock {\em \apj}, {\bf 626}, 350, 2005.

\bibitem[{Ott} et~al.(2006{\natexlab{a}}){Ott}, {Burrows}, {Thompson}, {Livne},
  and {Walder}]{ott:06spin}
C.~D. {Ott}, A.~{Burrows}, T.~A. {Thompson}, E.~{Livne}, and R.~{Walder}.
\newblock {\em Astrophys. J. Suppl. Ser.}, {\bf 164}, 130, 2006.

\bibitem[{Woosley} and {Heger}(2006)]{woosley:06}
S.~E. {Woosley} and A.~{Heger}.
\newblock {\em Astrophys. J.}, {\bf 637}, 914, 2006.

\bibitem[{Fryer} and {Heger}(2005)]{fryer:05}
C.~L. {Fryer} and A.~{Heger}.
\newblock {\em \apj}, {\bf 623}, 302, 2005.

\bibitem[{Burrows} et~al.(2007{\natexlab{c}}){Burrows}, {Dessart}, {Livne},
  {Ott}, and {Murphy}]{burrows:07b}
A.~{Burrows}, L.~{Dessart}, E.~{Livne}, C.~D. {Ott}, and J.~{Murphy}.
\newblock {\em \apj}, {\bf 664}, 416, 2007.

\bibitem[{LeBlanc} and {Wilson}(1970)]{leblanc:70}
J.~M. {LeBlanc} and J.~R. {Wilson}.
\newblock {\em \apj}, {\bf 161}, 541, 1970.

\bibitem[{Meier} et~al.(1976){Meier}, {Epstein}, {Arnett}, and
  {Schramm}]{meier:76}
D.~L. {Meier}, R.~I. {Epstein}, W.~D. {Arnett}, and D.~N. {Schramm}.
\newblock {\em \apj}, {\bf 204}, 869, 1976.

\bibitem[{Bisnovatyi-Kogan} et~al.(1976){Bisnovatyi-Kogan}, {Popov}, and
  {Samokhin}]{bisno:76}
G.~S. {Bisnovatyi-Kogan}, I.~P. {Popov}, and A.~A. {Samokhin}.
\newblock {\em \apss}, {\bf 41}, 287, 1976.

\bibitem[{Symbalisty}(1984)]{symbalisty:84}
E.~M.~D. {Symbalisty}.
\newblock {\em \apj}, {\bf 285}, 729, 1984.

\bibitem[{Wheeler} et~al.(2002){Wheeler}, {Meier}, and {Wilson}]{wheeler:02}
J.~C. {Wheeler}, D.~L. {Meier}, and J.~R. {Wilson}.
\newblock {\em \apj}, {\bf 568}, 807, 2002.

\bibitem[{Akiyama} et~al.(2003){Akiyama}, {Wheeler}, {Meier}, and
  {Lichtenstadt}]{akiyama:03}
S.~{Akiyama}, J.~C. {Wheeler}, D.~L. {Meier}, and I.~{Lichtenstadt}.
\newblock {\em \apj}, {\bf 584}, 954, 2003.

\bibitem[{Shibata} et~al.(2006){Shibata}, {Liu}, {Shapiro}, and
  {Stephens}]{shibata:06}
M.~{Shibata}, Y.~T. {Liu}, S.~L. {Shapiro}, and B.~C. {Stephens}.
\newblock {\em \prd}, {\bf 74(10)}, 104026, 2006.

\bibitem[{Sawai} et~al.(2008){Sawai}, {Kotake}, and {Yamada}]{sawai:08}
H.~{Sawai}, K.~{Kotake}, and S.~{Yamada}.
\newblock {\em \apj}, {\bf 672}, 465, 2008.

\bibitem[{Takiwaki} et~al.(2009){Takiwaki}, {Kotake}, and {Sato}]{takiwaki:09}
T.~{Takiwaki}, K.~{Kotake}, and K.~{Sato}.
\newblock {\em \apj}, {\bf 691}, 1360, 2009.

\bibitem[{Woosley} and {Bloom}(2006)]{wb:06}
S.~E. {Woosley} and J.~S. {Bloom}.
\newblock {\em Ann. Rev. Astron. Astrophys.}, {\bf 44}, 507, 2006.

\bibitem[{Dessart} et~al.(2008){Dessart}, {Burrows}, {Livne}, and
  {Ott}]{dessart:08a}
L.~{Dessart}, A.~{Burrows}, E.~{Livne}, and C.~D. {Ott}.
\newblock {\em \apjl}, {\bf 673}, L43, 2008.

\bibitem[{Cerd{\'a}-Dur{\'a}n} et~al.(2007){Cerd{\'a}-Dur{\'a}n}, {Font}, and
  {Dimmelmeier}]{cerda:07}
P.~{Cerd{\'a}-Dur{\'a}n}, J.~A. {Font}, and H.~{Dimmelmeier}.
\newblock {\em \aap}, {\bf 474}, 169, 2007.

\bibitem[{Obergaulinger} et~al.(2009){Obergaulinger}, {Cerd{\'a}-Dur{\'a}n},
  {M{\"u}ller}, and {Aloy}]{obergaulinger:09}
M.~{Obergaulinger}, P.~{Cerd{\'a}-Dur{\'a}n}, E.~{M{\"u}ller}, and M.~A.
  {Aloy}.
\newblock {\em \aap}, {\bf 498}, 241, 2009.

\bibitem[{Akiyama} and {Wheeler}(2005)]{akiyama:05}
S.~{Akiyama} and J.~C. {Wheeler}.
\newblock {\em \apj}, {\bf 629}, 414, 2005.

\bibitem[{Ott} et~al.(2007{\natexlab{a}}){Ott}, {Dimmelmeier}, {Marek},
  {Janka}, {Hawke}, {Zink}, and {Schnetter}]{ott:07prl}
C.~D. {Ott}, H.~{Dimmelmeier}, A.~{Marek}, H.-T. {Janka}, I.~{Hawke},
  B.~{Zink}, and E.~{Schnetter}.
\newblock {\em \prl}, {\bf 98}, 261101, 2007.

\bibitem[{Ott} et~al.(2007{\natexlab{b}}){Ott}, {Dimmelmeier}, {Marek},
  {Janka}, {Zink}, {Hawke}, and {Schnetter}]{ott:07cqg}
C.~D. {Ott}, H.~{Dimmelmeier}, A.~{Marek}, H.-T. {Janka}, B.~{Zink},
  I.~{Hawke}, and E.~{Schnetter}.
\newblock {\em Class. Quant. Grav.}, {\bf 24}, 139, 2007.

\bibitem[{Dimmelmeier} et~al.(2007){Dimmelmeier}, {Ott}, {Janka}, {Marek}, and
  {M{\"u}ller}]{dimmelmeier:07}
H.~{Dimmelmeier}, C.~D. {Ott}, H.-T. {Janka}, A.~{Marek}, and E.~{M{\"u}ller}.
\newblock {\em \prl}, {\bf 98(25)}, 251101, 2007.

\bibitem[{Dimmelmeier} et~al.(2008){Dimmelmeier}, {Ott}, {Marek}, and
  {Janka}]{dimmelmeier:08}
H.~{Dimmelmeier}, C.~D. {Ott}, A.~{Marek}, and H.-T. {Janka}.
\newblock {\em \prd}, {\bf 78(6)}, 064056, 2008.

\bibitem[{Watts} et~al.(2005){Watts}, {Andersson}, and {Jones}]{watts:05}
A.~L. {Watts}, N.~{Andersson}, and D.~I. {Jones}.
\newblock {\em \apjl}, {\bf 618}, L37, 2005.

\bibitem[{Centrella} et~al.(2001){Centrella}, {New}, {Lowe}, and
  {Brown}]{centrella:01}
J.~M. {Centrella}, K.~C.~B. {New}, L.~L. {Lowe}, and J.~D. {Brown}.
\newblock {\em \apjl}, {\bf 550}, L193, 2001.

\bibitem[{Ott} et~al.(2005){Ott}, {Ou}, {Tohline}, and {Burrows}]{rotinst:05}
C.~D. {Ott}, S.~{Ou}, J.~E. {Tohline}, and A.~{Burrows}.
\newblock {\em Astrophys. J.}, {\bf 625}, L119, 2005.

\bibitem[{Scheidegger} et~al.(2008){Scheidegger}, {Fischer}, {Whitehouse}, and
  {Liebend{\"o}rfer}]{scheidegger:08}
S.~{Scheidegger}, T.~{Fischer}, S.~C. {Whitehouse}, and M.~{Liebend{\"o}rfer}.
\newblock {\em \aap}, {\bf 490}, 231, 2008.

\bibitem[{Ott} et~al.(2006{\natexlab{b}}){Ott}, {Burrows}, {Dessart}, and
  {Livne}]{ott:06prl}
C.~D. {Ott}, A.~{Burrows}, L.~{Dessart}, and E.~{Livne}.
\newblock {\em \prl}, {\bf 96(20)}, 201102, 2006.

\bibitem[{Weinberg} and {Quataert}(2008)]{weinberg:08}
N.~N. {Weinberg} and E.~{Quataert}.
\newblock {\em \mnras}, {\bf 387}, L64, 2008.

\bibitem[ET()]{ET}
URL \url{http://www.et-gw.eu}.
\newblock {Einstein Telescope}.

\bibitem[{Ando} et~al.(2005){Ando}, Beacom, and Y\"uksel]{ando:05}
S.~{Ando}, F.~Beacom, and H.~Y\"uksel.
\newblock {\em Phys. Rev. Lett.}, {\bf 95}, 171101, 2005.

\end{thebibliography}
\end{document}